# Effects of Rashba-spin-orbit coupling on superconducting boron-doped nanocrystalline diamond films: evidence of interfacial triplet superconductivity

**Somnath Bhattacharyya** [1,2]*, **Davie Mtsuko**[1], **Christopher Allen**[3] **and Christopher Coleman**[1]

[1] Nano-Scale Transport Physics Laboratory, School of Physics, and DST/NRF Centre of Excellence in Strong Materials, University of the Witwatersrand, Private Bag 3, WITS 2050, Johannesburg, South Africa
[2] National University of Science and Technology MISiS, Moscow 119049, Leninsky pr. 4, Russia
[3] Department of Material Science, University of Oxford, OX1 3PU Oxford, UK

E-mail: Somnath.Bhattacharyya@wits.ac.za

## Abstract

Among the many remarkable properties of diamond, the ability to superconduct when heavily doped with boron has attracted much interest in the carbon community. When considering the nanocrystalline boron doped system, the reduced dimensionality and confinement effects have led to several intriguing observations most notably, signatures of a mixed superconducting phase. Here we present ultra-high-resolution transmission electron microscopy imaging of the grain boundary and demonstrate how the complex microstructure leads to enhanced carrier correlations. We observe hallmark features of spin-orbit coupling (SOC) manifested as the weak anti-localization effect. The enhanced SOC is believed to result from a combination of inversion symmetry breaking at the grain boundary interfaces along with antisymmetric confinement potential between grains, inducing a Rashba-type SOC. From a pronounced zero bias peak in the differential conductance, we demonstrate signatures of a triplet component believed to result from spin mixing caused by tunneling of singlet Cooper pairs through such Rashba-SOC grain boundary junctions.

## 1. Introduction

The search for unconventional superconducting systems has attracted great interest due to their potential applications in topological quantum computing[1]. In particular, there has been a strong focus on investigating potential systems that allow for the observation of triplet superconductivity[2] as it can be used as a platform for studying topological insulators[3]. The past decade has seen the much research directed at engineering heterostructures where triplet formation can be induced by proximity coupling of s-wave superconducting materials to ferromagnetic interfaces[4,5]. However, there also exists a growing list of materials that claimed intrinsic triplet superconductivity[6] including uranium-based compounds (UPt$_3$[7] and UTe$_2$[8]) and more recently the Fe-based superconducting systems[9,10,11] based on their special structures and internal properties. Although this list shows a diverse set of materials, one commonality is the presence of a complex structure, mostly layered, where internal interfaces with symmetry breaking properties allow for the formation of a triplet component. Theoretical studies have shown that a crucial ingredient for forming a mixed singlet-triplet superconducting condensate is the effect of spin-orbit coupling (SOC) on superconductivity[12-14]. This is due to extrinsic breaking of the symmetries of the pair potential as they pass through a region with significant SOC. A description of this effect involved the Hamiltonian term put forward by Gor'kov and Rashba[13] as $H_{SO} = \alpha(\boldsymbol{\sigma} \times \boldsymbol{p}) \cdot \boldsymbol{n}$, where $\boldsymbol{p}$ is the 2D quasi-momentum, $\boldsymbol{\sigma}$ are the Pauli matrices and $\boldsymbol{n}$ is the unit vector normal to the surface. The spin orbit interaction ensures that electrons in the plane ($\perp \boldsymbol{n}$) will have spin aligned perpendicular to the momentum $\boldsymbol{p}$. Although this effect was first described for purely 2D systems, subsequent variations have been formulated for interfacial systems of superconducting materials by Edelstein as interface SO [14]:

$$H_{SO} = \alpha(\boldsymbol{p} \times \boldsymbol{c}) \cdot \boldsymbol{\sigma}\delta(\boldsymbol{c} \cdot \boldsymbol{r}), \quad (1)$$

Here $\boldsymbol{c}$ is one of the two nonequivalent normal unit vectors and the $\boldsymbol{\delta}$ function describes the interface potential with a position vector ($\boldsymbol{r}$). Transport in systems such as granular superconductors relies on tunneling through the grain boundary interface. Electrons moving between grains results in a double electric layer on the same scale of the screening length. Charge carriers conducting through this layer would be subjected to a nonequivalent electric field (defined by the interface potential as a step function), and thus a spin orbit coupling of the Rashba-type arises. The non-equivalence of the two normals ensures that inversion symmetry along the boundary is broken and therefore the spin component of Cooper pairs traversing the interface will no longer be well defined, allowing for mixed states to occur, essentially converting singlet Cooper pairs to the short ranged triplet. This model is applicable to the nanocrystalline diamond films where grain boundaries form sharp interfacial potential (see Fig. 1).

Although it was believed that SOC is weak in carbon, there is much evidence suggesting that it can still have significant consequences on the electronic structure of surface and interface of various diamond systems[15-20]. Not only have hydrogen terminated and liquid-ion gated diamond surfaces recently demonstrated a strong and highly tunable SOC[15,16] but the boron acceptor itself is well known to harbor several interesting phenomena due to pronounced SOC[17]. Furthermore, in nanocrystalline diamond films, the reduced dimensionality and antisymmetric confining potential between grains can lead to an enhanced Coulomb repulsion. This coupled to the breaking of translational symmetry at the crystal boundary is expected to have major consequences on the properties of the films. In fact, the properties of the grain boundary in nanocrystalline boron doped diamond films have been linked to several unusual features of this material, including observations of the so called anomalous bosonic insulator[18,19,20] and demonstration of ferromagnetism[21]. Even more notable is the variation between scanning tunnelling microscopy (STM) measurements of this system; although initial STM studies have shown predominantly s-wave singlet pairing[22], more recent investigations have demonstrated a range of unexpected findings such as anisotropic tunnelling spectra[23], inhomogeneous gap structure across grains[24], finite energy resonances within the vortex core[25] and most notably, evidence of a mixture of Bardeen–Cooper–Schrieffer (BCS) and non-BCS pairing[26]. Even more intriguing is the fact that superconducting boron doped diamond films have already demonstrated several of the crucial features of a triplet superconducting system including an unusually high and anisotropic upper critical field[27] and a temperature independent nuclear magnetic resonance (NMR) Knight shift[28,29,30]. Furthermore, signatures of a ferromagnetic grain boundary[21] have been claimed for this system however the under lying mechanism for triplet pair formation was not well established. Such findings again support our claim of the triplet superconducting state however we believe the origin to be more likely a result of the grain boundary interface where a Rashba-type SOC can occur as described in Ref. 13 and 14.

In this work we demonstrate the existence of SOC through the observation of the weak anti-localization (WAL) effect in the low field magnetoresistance. By probing the charge transport through the granular films, we observe a pronounced zero bias peak which is robust to applied magnetic field, another hallmark features of a triplet superconducting state. These features as well as previous studies are reconciled through ultra-high resolution transmission electron microscopy (UHRTEM) investigations of the grain boundaries which show sharp crystal twinning and layered staking faults (interfacial region between grains) which we believe to be the origin of the observed enhanced SOC of the Rashba-type and thus ensuring a spin active boundary and the mixing of singlet and triplet pairing.

**2. Results and Discussion**

2.1. Microstructure

The microstructure of poly and nanocrystalline diamond films has been extensively studied with much of the research directed towards the properties of the grain boundaries[31]. It has been established that the surface termination of the diamond grain results in the formation of an extended $\pi^*$ orbital configuration due to hybridization of dangling bonds[32]. It has also been shown that heightened stress at such grain boundary regions leads to surface states through a modification of the electronic energy, thus forming frontier electronic orbitals[33]. Such findings are used to explain the many reports on the grain boundary conduction of nanocrystalline diamond systems[34]. Furthermore, there is also much evidence suggesting that carbon can form complex superlattice-like structures through alternating $sp^2$ and $sp^3$ bonding[35,36], thus supporting the interpretation as an array of S-I-S junctions (figure 1 h) This becomes particularly important when evaluating the superconducting boron doped diamond system as the granular nature ensures the occurrence of tunnelling and thus enhanced Coulomb repulsion between grains[36], leading to Cooper pair confinement[18,19,20]. All these studies indicate that the electronic transport in granular diamond systems is greatly dependent on the grain boundary region.

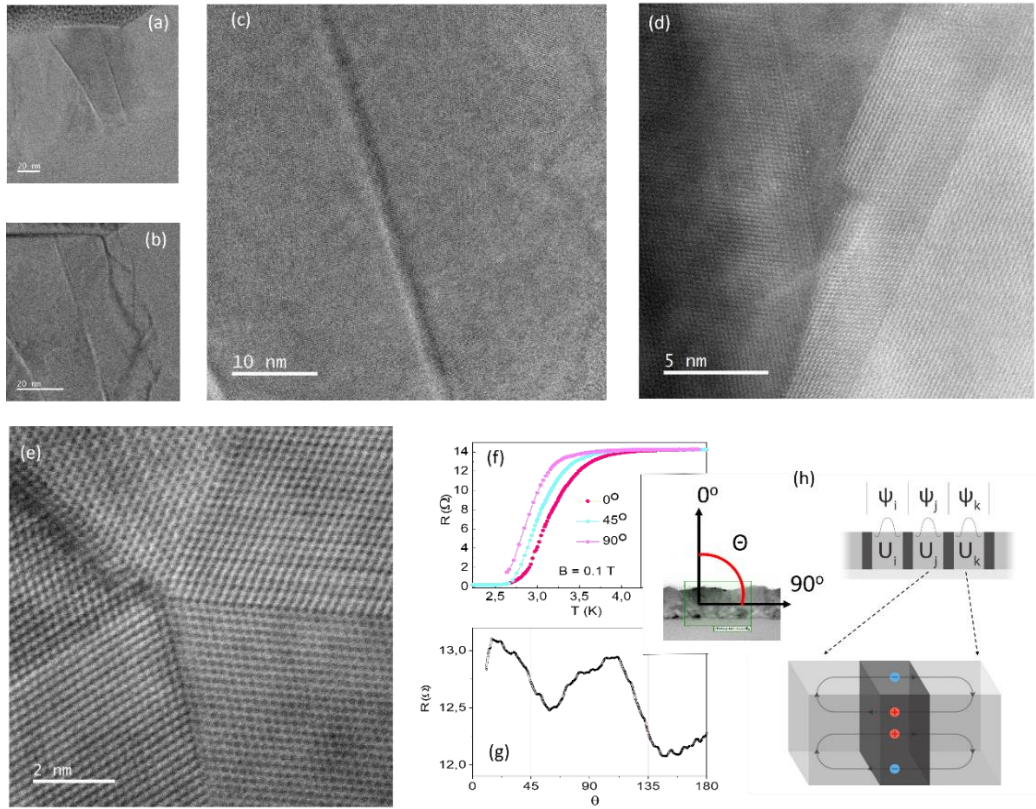

Fig 1. Microstructure analysis of CVD grown diamond films oriented close to the [110] zone axis. a) HAADF-STEM imaging of the nanocrystalline granular thin films. a) The thin films (~100 nm) show columnar growth where the length of diamond grain spans the entire film thickness. b) At higher magnification, the grain boundaries become more prominent with clear indications of stacking faults and defect planes between the individual grains forming distinct surfaces or 2D planes. c) When imaging the grain boundary region it is apparent that the interface region between grains can be as thick as a few nanometres and d) that individual diamond lattices can join at highly acute angles, ensuring strong translational symmetry breaking at such interfaces e) At certain points the diamond grains present pronounced crystal twinning as well, these are however located within individual grains. f) The transition temperature is also highly dependent on the angle of an applied field, as the angle of applied field increases with respect to the normal of the film, the critical temperature decreases, g) The resistance of the film shows a strong angle dependence when rotated in a magnetic field of 0.1 T, indicating that grain boundaries significantly affect the transport properties (data obtained at 3.25 K) h) this anomalous anisotropic angle dependent magnetoresistance has been observed in this system before and argued to result from confinement effects of the superconducting order parameter where a confining potential ($U_i$) restricts to Cooper pair wavefunctions ($\psi_i$) to individual grains, essentially quantum wells separated by an insulating boundary, such an interpretation therefore requires that mid gap bound states (multiple Andreev reflections) mediate the tunnelling between grains as in the case of an S-I-S junction arrays and that surface states can occur at such interfaces.

In figure 1 (a-d), high-angle annular dark field (HAADF) STEM imaging of the boron doped diamond has been used to image the nanocrystalline diamond films. As shown in figure 1 a, the individual grains are predominantly columnar, where single grains can extend throughout the entire film thickness. At higher magnification, the boundary regions become more prominent (figure 1 b). In figure 1 c, we demonstrate that the system is composed of highly ordered crystalline grains separated from each other by a sharp boundary region. The lattice miss match between such regions can be large, leading to stark crystal symmetry breaking, providing evidence of an interfacial (2D) structure between grains. In figure 1 d and e, it is shown that within the grains twinning frequently occurs. In order to investigate the effects of the microstructure on the transport, angle dependent magnetoresistance is measured. As shown in figure 1 (f and g) the resistance of the granular diamond films has a pronounced dependence on the angle of the applied magnetic field (rotation angle $\Theta = 0$ when applied normal to the film), not only does the resistance value oscillate when rotating the sample in a magnetic field of 0.1 T, but also demonstrates a shifting in the critical temperature (figure 1 (f)). Such features have been widely investigated in superconducting spin valve systems and are a result of the magnetization non-collinearity as Cooper pairs tunnel into a spin active domain. Similar features have recently been reported[27] and there also related to confinement along the grain boundaries.

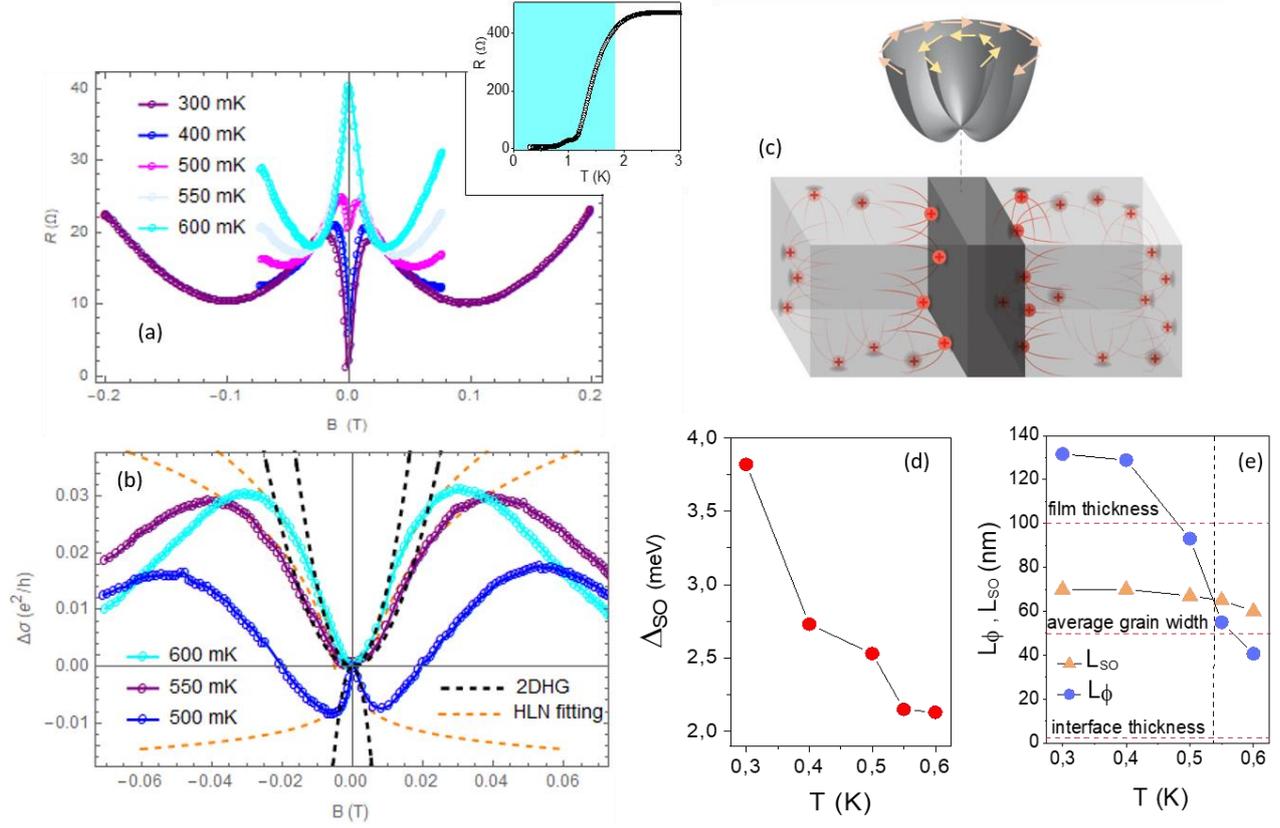

Fig 2. a) Magnetoresistance along the fluctuation regime, the isotherms indicate a complex series of transitions at low field far below the upper critical point (~ 1.5 T) showing a pronounced positive MR at lowest temperature indicative of a weak anti-localization effect. Upon increasing the temperature the slope is inverted to a negative MR, marking a transition to the weak localization regime, the temperature range where data was obtained is indicated in the inset b) the MR is re-plotted as magnetoconductance and fitted to the HLN formula as well as modified equation 2, which takes into account the corrections to conductivity with SOC, the formula fits well up to approximately 0.03 T, where an inversion to positive resistance is seen c) differences in the density of states between diamond grains can lead to charging effects and an effective antisymmetric confining potential (EC), this coupled to inversion symmetry breaking at the diamond grain boundaries can lead to a Rashba-type spin-orbit coupling where the spin bands are split at the surface. d) and e) From the fitting the temperature dependence of the Rashba spin orbit splitting, coherence length $L_\varphi$ and spin coherence length $L_{SO}$ are determined taking on values close to previous reports for diamond surfaces. These values are compared to various scales of the diamond films such as interface thickness, average grain size and film thickness as determined from HRTEM.

The observation here indicates not only the interference between the applied magnetic field and spin active interfaces but also demonstrate the intimate link between the microstructure and charge carrier transport. As argued before, when considering the confinement effects of the grain boundary region, the system behaves as an array of isolated grains where the Cooper pair wave function is restricted to individual grains by a confining potential related to the asymmetric charging that occurs between grains due to variation in the density of states (see figure 1 h). This charging effect along with the sharp inversion symmetry breaking at the grain boundary ensures that charge carriers conducting along the boundary region will be influenced by SOC.

2.2 Magnetoresistance
The superconducting phase transition shows a broad temperature dependence, this is known for granular superconducting systems as the coupling between grains is reduced due to enhanced Coulomb repulsion leading to a more resistive transition than what is observed in clean single crystal superconducting materials. In order to shed light on the microscopic character of the normal state charge carriers responsible for this broadening of the transition temperature, magnetoresistance (MR) measurements are conducted at temperatures near the critical point. As shown in figure 2(a) (figure 2b), the low field magnetoresistance (magnetoconductance) shows a complex series of transitions, at temperatures within the superconducting (zero resistance) phase; (i) the applied field firstly breaks the superconducting state leading to an initial increase in resistance which (ii) peaks then decreases steeply before (iii) reaching a temperature dependent minimum. By further increasing the field the system then only shows increasing resistance in accordance to field induced pair breaking. When increasing the temperature, the zero-field resistance can be seen to invert, and finally demonstrates a sharp peak centred at zero magnetic field.

Such non-monotonic resistance features are not at all expected for superconducting systems showing purely field induced Cooper pair breaking but are rather more likely to result from WAL due to the presence of spin orbit coupling, which has been widely observed in diamond interfaces[38,39,40]. The observation of the WAL effect is significant as it is a hallmark feature of spin orbit coupling in a 2D system, and considering the confinement as well as charging effects intrinsic to the nanocrystals that compose the film we attribute this observation to the interfacial region between individual grains (shown in figure 1 (a)-(c)). To date, a large number of reports exist on the observation of the WAL effect in diamond surfaces[38,39,40], where this effect has been reported to exist due to a Rashba spin-orbit coupling (RSOC) that results from the structural symmetry breaking at the diamond edge and unbalanced confinement potential, both of which are intrinsic properties of the grain boundary in our nanocrystalline films this if schematically indicated in figure 2(c).

In order to deeper investigate this phenomenon, we utilize conventional analysis relying on the Hikami-Larkin-Nagaoka (HLN) formalism[41]:

$$\Delta\sigma = \sigma(B) - \sigma(0) = \frac{\alpha e^2}{2\pi^2 \hbar}[\ln\left(\frac{\hbar}{4BeL_\varphi^2}\right) - \psi(\frac{1}{2}+\frac{\hbar}{4BeL_\varphi^2})] \quad (1)$$

as well as its modified version which has been specifically applied to hole type carriers in diamond surfaces[42]:

$$\Delta\sigma = -\frac{e^2}{2\pi^2\hbar}\left[\psi\left(\frac{1}{2}+\frac{B_\varphi+B_{SO}}{B}\right)\right]+\frac{1}{2}\psi\left(\frac{1}{2}+\frac{B_\varphi+2B_{SO}}{B}\right)-\frac{1}{2}\psi\left(\frac{1}{2}+\frac{B_\varphi}{B}\right) -\ln\left(\frac{B_\varphi+B_{SO}}{B}\right)-\frac{1}{2}\left(\frac{B_\varphi+2B_{SO}}{B}\right)+\ln\left(\frac{B_\varphi}{B}\right) \quad (2)$$

In the first equation the only fitting parameters are the pre-factor $\alpha$ and $L_\varphi$, the coherence length whereas the second equation defines the phase coherence ($B_\varphi$) and spin coherence fields ($B_{SO}$). The pre-factor $\alpha$ takes on values of between 0.5 and 1 and allows us to determine an overall validity of our interpretation of the data. As shown in figure 2 (b) the second equation captures the low field features of the magnetoconductance where temperature dependence of the fitting parameters are plotted in figure 2 (d) and (e). Following the previous reports of WAL in diamond systems, we assume a Rashba splitting[38] and from the fit we can extract the temperature dependence of the phase and spin coherence lengths which are given by $L_\varphi = (\hbar/4eB_\varphi)^{0.5}$ and $L_{SO} = (\hbar/4eB_{SO})^{0.5}$, respectively. As shown in figure 2 (d), the phase coherence length shows a much stronger temperature dependence when compared to the spin coherence length, which is nearly temperature independent in the evaluated temperature range.

One additional observation is that the crossover of the WAL to a WL observed in the magnetoresistance corresponds to the temperature where the spin coherence length exceeds the phase coherence length. This is an indication of the spin precession destroying the phase coherence of the hole carriers[38,43,44,45]. It is also possible to interpret this crossover in the magnetoresistance as evidence of the system moving from a regime dominated by surface conduction (or interfacial grain boundary transport) to one where charge carriers of the bulk dominate at elevated temperature. This interpretation is supported by the recent report from Zhang et al.[27] where the anomalous angle dependent magnetoresistance demonstrated carrier localization at the grain boundary. From the extraction of the spin coherence field, it is also possible to determine the spin relaxation time $\tau_{SO} = \hbar/4eDB_{SO}$, (where D is the diffusion constant and related to the hole carrier density). Using this value, we determine the spin-orbit splitting given by the following relationship: $\Delta_{SO} = \hbar/(2\tau\tau_{SO})^{0.5}$. The temperature dependence of this spin-orbit splitting is plotted in figure 2 (e) and although the values are smaller than what has been reported for gated diamond surfaces[39], the values are within the range reported for other quantum well structures[46,47]. Although this is the first report of the observation of the WAL to WL crossover in nanocrystalline diamond films, the data does neatly follow the expected theory and yields realistic parameters in accordance to what has been cited in literature[39,46,47]. These findings thus further establish the significance of the diamond grain boundary in understanding the transport properties of this system. In particular it highlights the possibility of RSOC as an alternative explanation for the origin of spin active grain interfaces as recently reported for hydrogenated granular diamond systems[21].

2.3 Zero bias conductance peak (ZBCP)

Although much work has been directed at investigating the order parameter of this system through STM measurements[22,23,24,25,26], up to now many anomalous features reported in these works have not been reconciled. Although STM imaging results in high resolution spectra with a well-defined theoretical backing, it is only a local probe and thus is not ideal when considering dynamical processes such as those present when Cooper pairs tunnel through the grain boundaries[48]. This is particularly important in the present system as inversion symmetry breaking at the boundary leading to SOC at grain interface are expected to allow for spin mixing and triplet formation and can show ZBCP features.

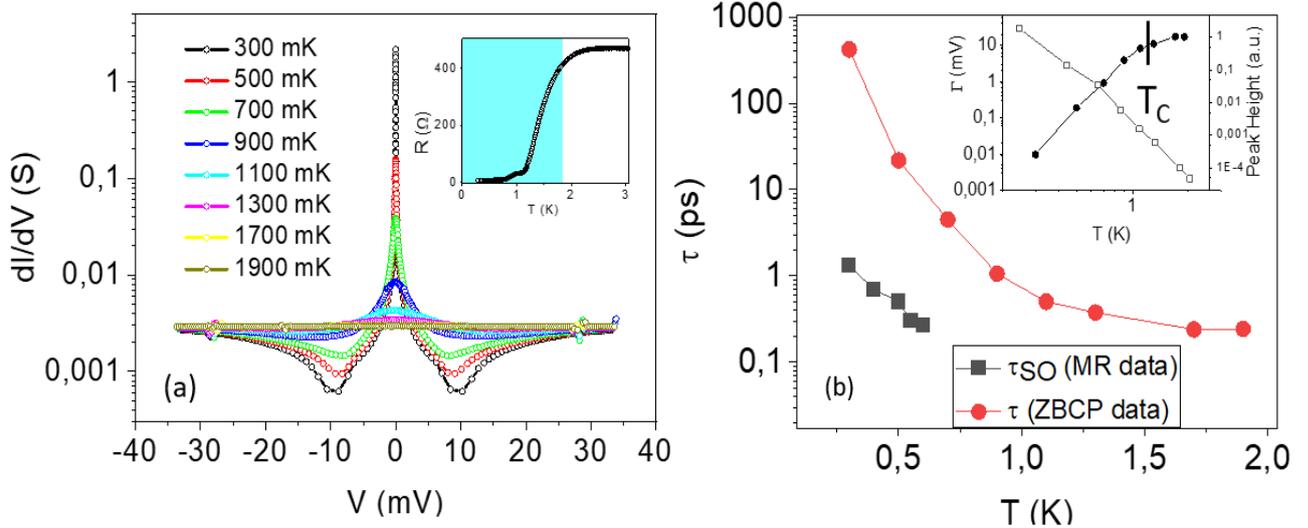

**FIG. 3 (a)** d*I/dV* as a function of voltage showing a pronounced zero bias conductance peak. The zero-energy peak is highly dependent on temperature and is only observed at temperatures below the superconductivity transition onset, as indicated in the inset. **(b)** From the FWHM we determine the estimated lifetime of the triplet mode, this shows an exponential dependence on time and saturates in the *ps* regime **(inset)** The peak height decreases exponentially with increasing temperature and shows a full width half maximum that increases exponentially upon increasing temperature till the critical temperature ($T_C$) is reached, then remains constant until no longer observed at higher temperatures.

In order to investigate spin-triplet and RSOC effects on superconductivity, we rely on transport four-probe voltage biased measurement through the diamond films. As shown in figure 3 (a), the differential conductance obtained by measuring current transport through the film does indeed show pronounced zero bias conductance peaks. As indicated in the inset of figure 4 (a), this ZBCP is only observed at temperature below the superconductivity onset and thus as such features have not been seen in single crystal samples or in the STM spectra of granular diamond samples, the most plausible explanation is that such features arise from the intergranular transport through the diamond films.

The formation of such ZBCP is well documented in superconducting systems that deviate from conventional s-wave singlet superconductivity and are a result of the formation of so-called mid gap bound states[49,50,51]. These bound states are caused by constructive interference between incident and Andreev reflected quasiparticles along a grain boundary, essentially surface states, and are intimately linked to the properties of the order parameter[51]. In the boron doped diamond films, these ZBCP have shown a pronounced dependence on temperature and applied field (fig. 4 (a) and fig. 5 (a), respectively). As the ZBCP is a type of resonance mode, by extracting the full width half maximum (FWHM) of the peak it is possible to determine the lifetime of the bound states that give rise to this feature. This is shown in figure 4 (b) where the extracted lifetime can be seen to decay exponentially with increasing temperature, reaching a minimum value within the *ps* regime. This is significant as from the magnetoresistance data the spin coherence lifetime assumed from the spin active interface can also found to be limited to the *ps* regime, an indication that the two are intimately linked. As can be seen in the inset figure 3 (b) the ZBCP height decreases with a power law scaling as temperature is increased and the FWHM shows a broadening with increased temperature (inset fig. 3 (b)). The peak is only observed at temperatures below the superconducting phase transition onset and quickly vanishes above this temperature.

Further support linking a triplet state to the observed ZBCP is the field dependence. Upon application of a magnetic field the peak height is reduced as shown in Fig 4 (a), indicating suppression of the superconducting phase. Another interesting observation is the robust nature of the zero-energy peak towards increasing magnetic field. As shown in figure 4(a) and (b) the zero-energy peak does not show any splitting with increasing field, a hallmark feature of spin triplet superconductivity[52]. Upon increasing the applied field strength, the peak height does however decrease until it is completely supressed at approximately 1.5 T, after which further increase of the field leads to a pronounced dip or V-shape in the differential conductance. Such a magnetic field dependent resistive behaviour has been well documented in various systems, either intrinsic or in heterostructures exhibiting triplet superconductivity[53,54,55,56].

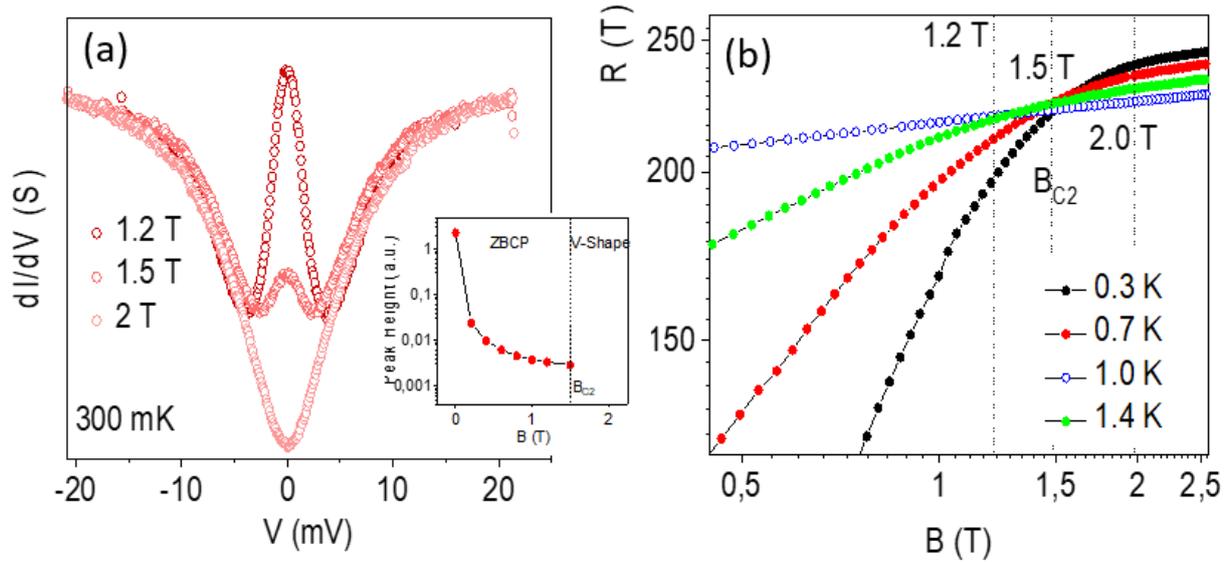

**Fig4 a)** The ZBCP does not show any indication of splitting up until it is completely suppressed and inverted to a V-shape dip, the field dependence of the peak height is given in the inset **(b)** The field strength where this inversion of the peak occurs marks the crossover ($B_{C2}$) from superconducting to insulating regime

A further observation is that the magnetic field strength where the peak is almost completely suppressed, corresponds to a crossing point in the high field magnetoresistance isotherms as indicated in figure 4(b). From these transport features it is apparent that the complex structure of the nanocrystalline diamond system, most notably the grain boundary interface, has a significant influence on the properties of the superconducting phase. Moreover, when considering the multilayer system with broken inversion symmetry, i.e. superlattice, along with possible RSOC as mentioned above, it is possible for the predicted spin mixing effect to occur. As there are several mechanisms describing the observation of ZBCP we further probe this feature by applying a magnetic field to rule out competing mechanisms, a thorough discussion is included in the supplementary information section.

Although the influence of Rashba induced spin polarization of Andreev bound states is an actively researched area, it is still not experimentally known to what extant Rashba SOC can influence these resonance modes and as shown in recent publications[57] the effect of spin obit coupled layers in superconducting heterostructures can greatly influence the transport. We have highlighted several anomalous features linked to the possibility of a triplet component in the boron doped diamond system, however further investigation will be required to fully understand the dynamics of this system

In this Letter analysis of these data firmly establishes the effect of Rashba-spin-orbit coupling on superconductivity in the boron doped diamond system. The SOC is expected to result from the broken inversion symmetry and asymmetric confining potential between crystal grains. We go on to show the spin coherence lifetime ($\tau_{SO}$) and triplet resonance lifetimes ($\tau$) are connected, indicating they share the same origin. The effect of the SOC on superconductivity in such as low dimensional superconducting system has huge implications for the nature of the superfluid condensate. As predicted by theory in such a system a single/triplet mixing occurs, this has been verified through the observation of pronounced zero bias conductance peak in the differential conductance which is robust to applied fields. These findings help us unify the previously reported variation in STM measurements and further establish the nanocrystalline boron doped diamond system as an unconventional superconductor with a triplet component.

**Acknowledgements**

SB is very thankful to Miloš Nesládek for the samples. He thanks A. Irzhak, S. Mukhin, S. Polyakov (from MISIS) for discussion and D. Churochkin for his valuable contribution to this project. SB would like to thank A. Kirkland (Oxofrd) for granting HRTEM beamline time and D. Wei and C. Hyunh (Carl Zeiss Microscopy) for lamella foil preparation. C.C. thanks W. Matthews for illustrative work. SB thanks CSIR-NLC, the URC Wits and National Research Foundation (SA) for the BRICS multilateral program for funding. SB also acknowledges financial support from the Ministry of Education and Science of the Russian Federation in the framework of the Increased Competitiveness Program of NUST « MISiS» (grant No. K3-2018-043).